\documentclass[aps,prl,twocolumn,showpacs,groupedaddress,amsmath,amssymb]{revtex4}
\usepackage{graphicx}
\usepackage{color}
\usepackage{bm}
\usepackage{latexsym}
\usepackage{ulem}

\begin{document}

\title{Critical States Embedded in the Continuum}
\author{M. Koirala$^1$, A. Yamilov$^1$, A. Basiri$^{2}$, Y. Bromberg$^3$,  H. Cao$^3$, T. Kottos$^{2}$}
\affiliation{$^1$Department of Physics, Missouri University of Science and Technology, Rolla, MO-65409, USA}
\affiliation{$^2$Department of Physics, Wesleyan University, Middletown, CT-06459, USA}
\affiliation{$^3$Department of Applied Physics, Yale University, New Haven CT-06520, USA}
\date{\today }

\begin{abstract}
We introduce a class of critical states which are embedded in the continuum (CSC) of one-dimensional optical 
waveguide array with one non-Hermitian defect. These states are at the verge of being fractal and have real 
propagation constant. They emerge at a phase transition which is driven by the imaginary refractive index of 
the defect waveguide and it is accompanied by a mode segregation which reveals analogies with the Dicke super
-radiance. Below this point the states are extended while above they evolve to exponentially localized modes. 
An addition of a background gain or loss can turn these localized states to bound states in the continuum. 
\end{abstract}
\maketitle

{\it Introduction -} A widespread preconception in quantum mechanics is that a finite potential well can support 
stationary solutions that generally fall into one of the following two categories: (a) Bound states that are square 
integrable and correspond to discrete eigenvalues that are below a well-defined continuum threshold; and (b) 
Extended states that are not normalizable and they are associated with energies that are distributed continuously 
above the continuum threshold~\cite{P93}. This generic picture has further implications. For example it was used 
by Mott~\cite{M67} in order to establish the presence of sharp mobility edges between localized and extended 
wavefunctions in disordered systems. Specifically it was argued that a degeneracy between a localized and an 
extended state would be fragile to any small perturbation which can convert the former into the latter. Nevertheless, 
von Neumann and Wigner succeeded to produce a counterintuitive example of a stationary solution which is square 
integrable and its energy lies above the continuum threshold~\cite{NW29}. Their approach, although conceptually 
simple, was based on reverse engineering i.e. they prescribed the state and then constructed the potential  that supports it. 
These, so-called, Bound States in the Continuum (BIC) are typically fragile to small perturbations which couples them 
to resonant states and the associated potential that supports them is usually complicated. At the same time, they can 
provide a pathway to confine various forms of waves like light~\cite{PPDHNSS11,WXKMTNSSK13,CVCOL13,HZLCJJS13}, 
acoustic, water waves~\cite{PE05}, and quantum~\cite{CSFSCC92} waves as much as to manipulate nonlinear phenomena 
in photonic devices for applications to biosensing and impurity detection~\cite{MBS08}.

Although most of the studies on the formation of BIC states have been limited to Hermitian systems there are, nevertheless, 
some investigations that address the same question in the framework of non-Hermitian wave mechanics~\cite{OPR03}. 
Along the same lines the investigation of defect modes in the framework of ${\cal PT}$-symmetric optics~\cite{ZGWL10,
RMBNOCP13,L14} has recently attracted some attention. Though the resulting defect states either are not BIC states as they 
emerge in the broken phase where the eigenfrequencies are complex (and thus the modes are non-stationary)~\cite{RMBNOCP13} 
or when they appear in the exact phase, and thus correspond to real frequencies, the resulting potential is complex and 
its realization is experimentally challenging~\cite{L14}.

In this paper we introduce a previously unnoticed class of critical states which are embedded in the continuum (CSC). 
We demonstrate their existence using a simple set-up consisting of $N$ coupled optical waveguides with one non-
Hermitian (with loss or gain) defective waveguide in the middle. Similarly to BIC they have real propagation constant; 
albeit their envelop resembles a fractal structure. Namely their inverse participation number ${\cal I}_2$ scales anomalously 
with the size of the system $N$ as
\begin{equation}
\label{eq1}
{\cal I}_2\equiv{\sum_n |\phi_n|^4\over \sum_n|\phi_n|^2}\sim {\log (N+1) \over (N+1)}
\end{equation}
Above $\phi_n$ is the wavefunction amplitude of the BIC state at the $n-$th waveguide. The CSC emerges in the middle 
of the band spectrum of the perfect array when the imaginary index of refraction of the defective waveguide $\epsilon_0^{(I)}$ 
becomes $|\epsilon_0^{(I)}|\geq2V$ where $V$ is the coupling constant between nearby waveguides. Below this value all 
modes of the array are extended while in the opposite limit the CSC becomes exponentially localized with an inverse localization 
length $\xi^{-1}=\ln[2V/({|\epsilon_0^{(I)}|-\sqrt{(\epsilon_0^{(I)})^2-4V^2}})]$  and the 
associated mode profile changes from non-exponential to exponential decay. The localization
-delocalization transition point is accompanied with a mode re-organization in the complex frequency plane which reveals 
many similarities with the Dicke super/sub radiance transition. Finally we can turn these exponentially localized modes to
BIC modes by adding a uniform loss (for gain defect) or gain (for lossy defect) in the array, thus realizing BIC states in a
simple non-Hermitian set-up.

{\it Physical set-up -} We consider a one-dimensional array of $N=2M+1$ weakly coupled single-mode optical waveguides. 
The light propagation along the $z$-axis is described by the standard coupled mode equations~\cite{CLS03}
\begin{equation}
\label{eq2}
i\lambdabar{\partial \psi_n(z)\over \partial z}+V\left(\psi_{n+1}(z)+\psi_{n-1}(z)\right) + \epsilon_n \psi_n(z)=0
\end{equation}
where $n=-M,\cdots,M$ is the waveguide number, $\psi_n(z)$ is the amplitude of the optical field envelope at distance $z$ 
in the $n$-th waveguide, $V$ is the coupling constant between nearby waveguides and $\lambdabar\equiv\lambda/2\pi$ 
where $\lambda$ is the optical wavelength in vacuum. The refractive index $\epsilon_n$ satisfies the relation $\epsilon_n=
\epsilon_0^{(R)}+i\epsilon_n^{(I)}\delta_{n,0}$ where we have assumed that a defect in the imaginary part of the dielectric 
constant is placed in the middle of the array at waveguide $n=0$. Below, without loss of generality, we will set $\epsilon_0^{(R)}=0$ 
for all waveguides \cite{note1}. Our results apply for both gain $\epsilon_0^{(I)}<0$ and lossy $\epsilon_0^{(I)}>0$  defects. 
Optical losses can be incorporated experimentally by depositing a thin film of absorbing material on top of the waveguide~
\cite{salamo}, or by introducing scattering loss in the waveguides \cite{szameit}. Optical amplification can be introduced by 
stimulated emission in gain material or parametric conversion in nonlinear material \cite{kipp}. 

Substitution in Eq.~(\ref{eq2}) of the form $\psi_n(z)=\phi_n^{(k)} \exp(-i\beta^{(k)} z/\lambdabar)$, where the propagation 
constant $\beta^{(k)}$ can be complex due to the non-Hermitian nature of our set-up, leads to the Floquet-Bloch (FB) eigenvalue 
problem 
\begin{equation}
\label{eq3}
\beta^{(k)} \phi_n^{(k)} =- V(\phi_{n+1}^{(k)}+\phi_{n-1}^{(k)}) - \epsilon_n\phi_n^{(k)}; \quad k=1,\cdots,N
\end{equation}
We want to investigate the changes in the structure of the FB modes and the parametric evolution of the propagation constants 
$\beta^{(k)}$ as the imaginary part of the optical potential $\epsilon_0^{(I)}$ increases.

{\it Mode segregation and Dicke super-radiance -} We begin by analyzing the parametric evolution of $\beta^{(k)}$'s as a function 
of the non-Hermiticity parameter $\epsilon_0^{(I)}$. We decompose the Hamiltonian $H_{nm}$ of Eq.~(\ref{eq3}) into a Hermitian 
part $(H_0)_{nm}=-V\delta_{n,m+1}-V\delta_{n,m-1}$ and a non-Hermitian part $\Gamma_{nm}=-i\epsilon_n^{(I)}\delta_{n,0}
\delta_{n,m}$ i.e. $H=H_0+\Gamma$. For $\epsilon_0^{(I)}=0$ the eigenvalues and eigenvectors of $H=H_0$ are $\beta^{(k)}=
-2V\cos(k\pi/(N+1))$ and $\phi_n^{(k)}=\sqrt{2/(N+1)} \sin\left[k\left(n\pi/(N+1)+\pi/2\right)\right]$. In the limit $N\rightarrow 
\infty$ the spectrum is continuous creating a band $\beta\in [-2V,2V]$ that supports radiating states. 

As $\epsilon_0^{(I)}$ increases from zero the propagation constants move into the complex plane. For small values of $\epsilon_0^{(I)}$ a perturbative 
picture is applicable and can explain satisfactorily the evolution of $\beta$'s. Using first order perturbation theory we get that $\beta^{(k)}\approx
\beta_0^{(k)}+\Gamma_{k,k}$ where $\Gamma_{k,k}\approx -i\epsilon_0^{(I)}/(N+1)$. When the matrix elements of the non-Hermitian part of $H$ 
become comparable with the mean level spacing $\Delta=2V/N$ of the eigenvalues of the Hermitian part $H_0$, the perturbation theory breaks down. 
This happens when $|\epsilon_{cr}^{(I)}|/(N+1) \sim \Delta$ which leads to the estimation $|\epsilon_{cr}^{(I)}|\sim 2V$. In the opposite limit of large 
$|\epsilon_0^{(I)}|$, $H_0$ can be treated as a perturbation to $\Gamma$. Due to its specific form, the non-Hermitian matrix $\Gamma$ has only one 
nonzero eigenvalue and thus, in the large $|\epsilon_0^{(I)}|$ limit, there is only one complex propagation constant corresponding to ${\cal R}e\left[
\beta_0^{(k=(N+1)/2)}\right]=0$, while all other modes will have zero imaginary component (to first order). The above considerations allow us to 
conclude that for $|\epsilon_0^{(I)}|\gg 2V$ a segregation of propagation constants in the complex plane occurs: Below this point all $\beta$'s get 
an imaginary part which increases in magnitude as $\sim -\epsilon_0^{(I)}/N$ while after that only one of them accumulates almost the whole imaginary 
part $\sim -\epsilon_0^{(I)}$ (independent of $N$) and the remaining $N-1$ approaches back to the real axis as $\sim -(2V)^2/(N\epsilon_0^{(I)})$. 
This segregation of propagating constants is the analogue of quantum optics Dicke super-radiance transition~\cite{D54} which was 
observed also in other frameworks~\cite{OPR03,SZ89,C12,KSKHSA12}. These predictions are confirmed by our numerical data (see Fig. \ref{fig1}). 

\begin{figure}[tbp]
\begin{center}
\includegraphics[width=3.3in]{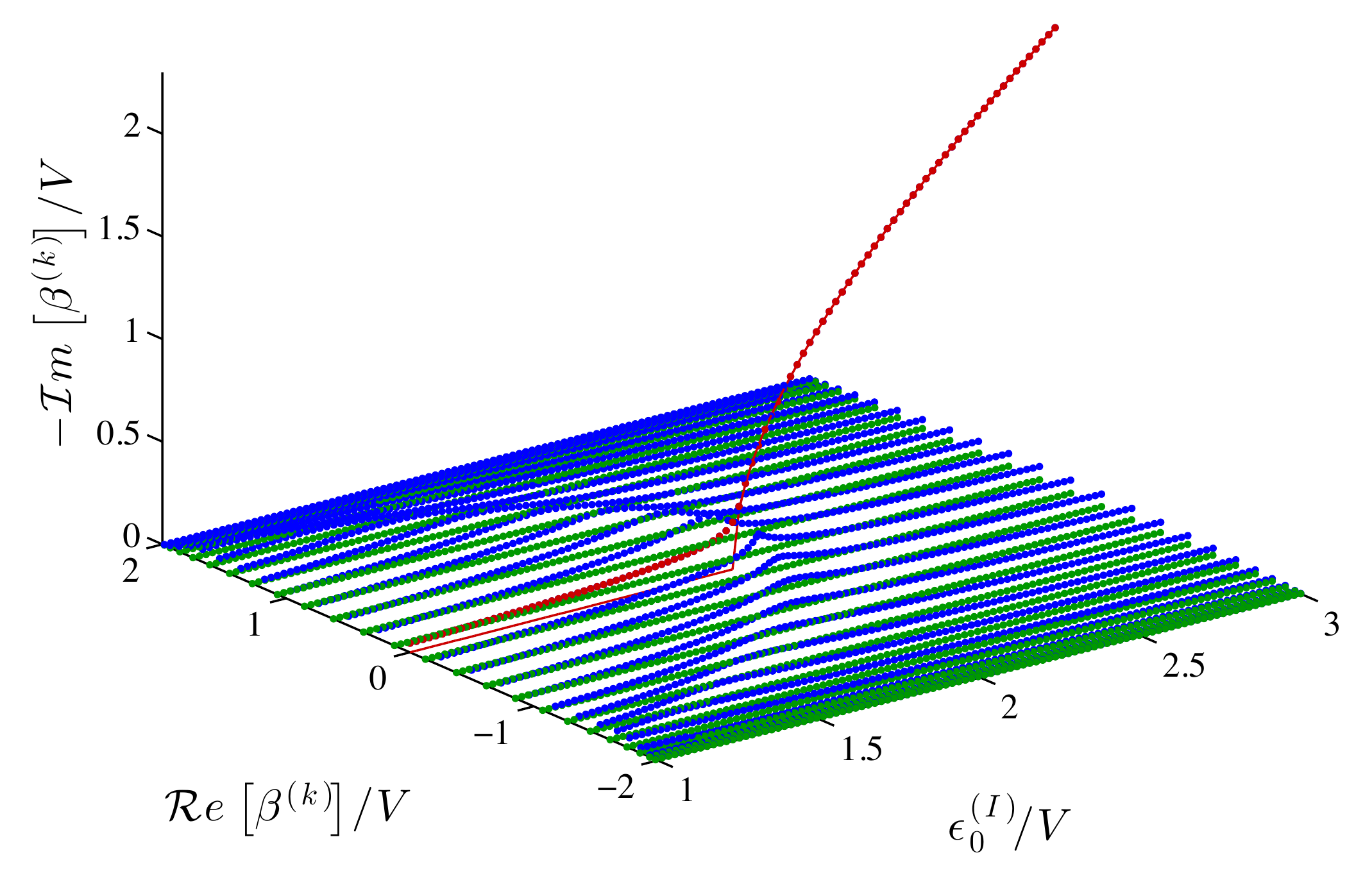}
\end{center}
\vskip -0.5cm
\caption{Parametric evolution of the propagation constants $\beta^{(k)}$ of an array of $N=49$ coupled waveguides with one dissipative 
($\epsilon_0^{(I)}>0$) defect in the middle, as a function of the non-Hermiticity $\epsilon_0^{(I)}$ of the defect. The super-radiance point 
is defined by the condition $\epsilon_0^{(I)}=\epsilon_{cr}^{(I)}\equiv+2V$ where the defect mode profile (shown as red dots) switches from 
non-exponential to exponential decay. Solid red line shows the asymptotic analytical result from Eq.~(\ref{eq5}). Similar behavior (not shown 
here) but with the $\beta$'s in the upper complex plane can be observed in the case of gain $\epsilon_0^{(I)}<0$ where $\epsilon_{cr}^{(I)}
=-2V$.}
\label{fig1}
\end{figure}

{\it Delocalization-localization transition and BIC- } Next we investigate the structure of the FB modes of the system Eq.~(\ref{eq3}) in the 
thermodynamic limit (N$\rightarrow$$\infty$) as $\epsilon_0^{(I)}$ crosses the threshold $\epsilon_{cr}^{(I)}$. In the case of real defect, 
we know that an infinitesimal value of it will lead to the creation of a localized mode (with a real-valued $\beta$ outside of the continuum 
$[-2V,2V]$ interval)~\cite{E06}. We want to find out if the same scenario is applicable in the case of imaginary defect. To this end we introduce 
the ansatz:
\begin{equation}
\label{eq4}
\phi_n= 
\left\{\begin{array}{cc}
A^{(+)} \exp(-n \Lambda) & {\rm for}\quad n\geq 0\\
A^{(-)} \exp( n  \Lambda)  & {\rm for} \quad n\leq 0
\end{array}
\right.
\end{equation}
Continuity requirement of the FB mode at $n=0$ leads to $A^{(+)}=A^{(-)}$. Furthermore, substituting the above ansatz in 
Eq.~(\ref{eq3}) for $n=0$ and $n=1$ and after some straightforward algebra we get that 
\begin{equation}
\label{eq5}
\beta=-s\sqrt{4V^2-(\epsilon_0^{(I)})^2} \quad {\rm and} \quad  \Lambda=-\ln\left({-\beta-i\epsilon_0^{(I)} \over 2V}\right),
\end{equation}
where $s\equiv\epsilon_0^{(I)}/|\epsilon_0^{(I)}|$ denotes the sign of the defect. From Eq.~(\ref{eq5}) we find that for 
$|\epsilon_0^{(I)}|<|\epsilon_{cr}^{(I)}|\equiv2V$ the corresponding propagation constant is real while the decay rate is 
$ \Lambda=-i\arctan\left(\epsilon_0^{(I)}/\beta\right)$ i.e. a simple phase. In other words the FB modes are extended. 
In the opposite limit of $|\epsilon_0^{(I)}|>|\epsilon_{cr}^{(I)}|$ the propagation constant becomes complex and the 
corresponding $\Lambda$ takes the form 
\begin{equation}
\label{eq6}
\Lambda=\ln\left({2V \over \left|\epsilon_0^{(I)}\right|-\sqrt{(\epsilon_0^{(I)})^2-4V^2}}\right) + i\,s\,{\pi\over 2}
\end{equation}
The corresponding inverse localization length is then defined as $\xi^{-1}\equiv{\cal R}e(\Lambda)$ indicating the existence 
of exponential localization. Therefore we find that a non-Hermitian defect - in contrast to a Hermitian one - induces a 
localization-delocalization transition at the Dicke super-radiance phase transition points $\epsilon_{cr}^{(I)}=s\times 2V$. We 
emphasize again that this phase transition and the creation of a localized mode occur for both signs of the non-Hermitian 
defect and can be induced for both lossy ($\epsilon_0^{(I)}>0$) and gain ($\epsilon_0^{(I)}<0$) defect. 

\begin{figure}[tbp]
\begin{center}
\includegraphics[width=3.3in]{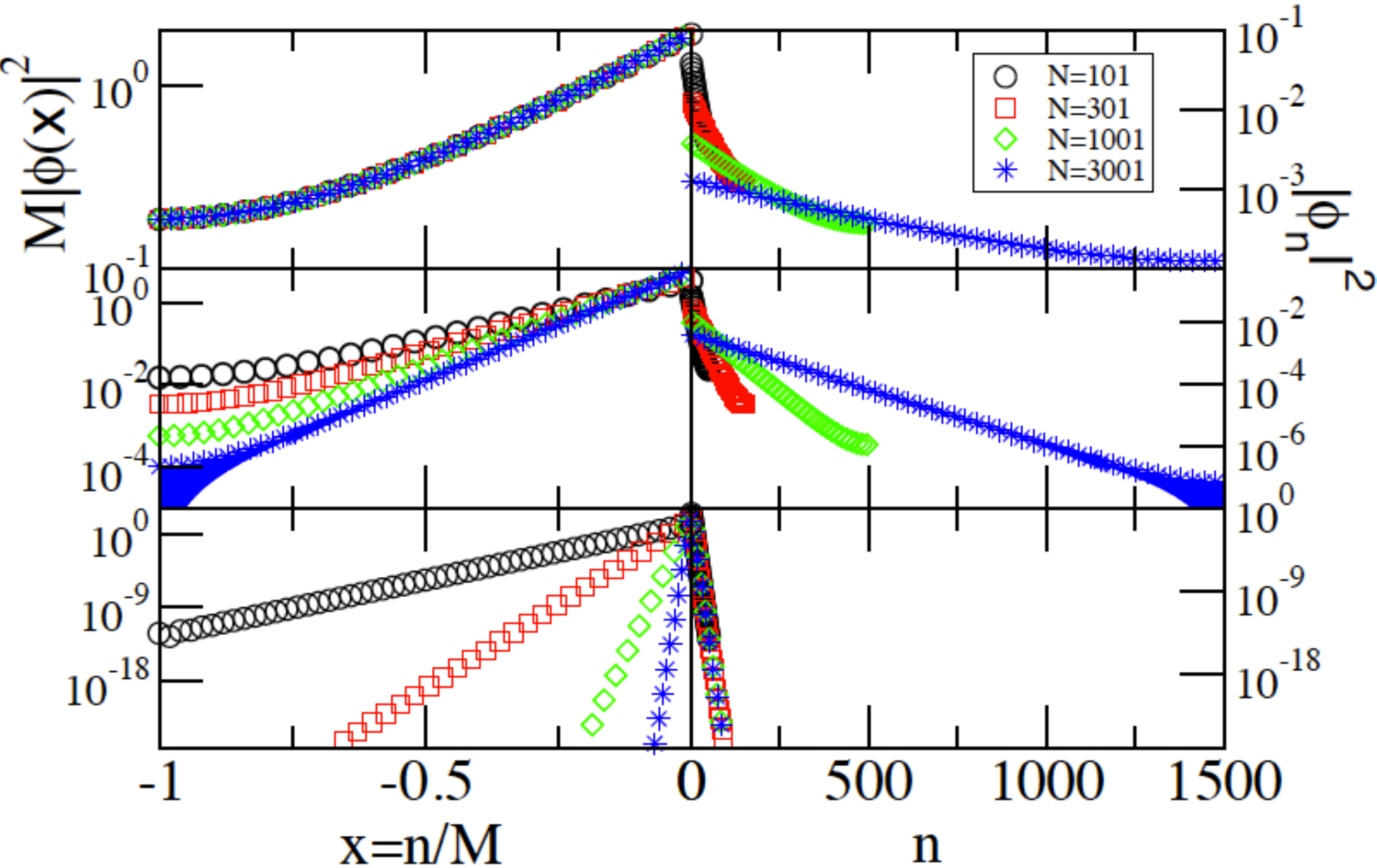}
\end{center}
\vskip -0.5cm
\caption{
Floquet-Bloch defect mode for various system sizes $N=2M+1$. Left panels report the left part ($n<0$) of these modes (the right 
part $n>0$ is the same) by employing the scaling $M\phi(x=n/M)$ while the right panels report the right part $n>0$ of these modes 
without any scaling. In the former representation an extended state is invariant under increase of the size of the system while in the 
latter, the scale invariance is demonstrated for localized modes. Three defect values of $\epsilon_0^{(I)}$ has been used: (upper) Below 
threshold $0<\epsilon_0^{(I)}<2V$ where the mode is delocalized; (middle) At threshold $\epsilon_0^{(I)}=2V$ where the mode is 
critical; (lower) Above threshold $\epsilon_0^{(I)}>2V$ where the mode is exponentially localized. Notice that in the case of critical 
modes (middle panels) }
\label{fig2}
\end{figure}

We have confirmed the theoretical analysis with numerical simulations. In Fig. \ref{fig2} we report the FB defect mode of our 
system Eq.~(\ref{eq3}) for three cases corresponding to (a) $0<\epsilon_0<2V$ (below threshold), (b) $\epsilon_0=2V$ (at threshold)
and (c) $\epsilon_0>2V$ (above threshold), and different system sizes. Note that although in the latter case the mode is localized 
in space, it is not qualified as a BIC since the corresponding propagation constant $\beta$ (see Eq. (\ref{eq5})) is imaginary and 
therefore the mode is non-stationary. Adding, however, a uniform gain (for lossy defect) $\beta$ or loss (for gain defect) 
$-\beta$ to the array can turn this state to a BIC with zero imaginary propagation constant. The latter case is experimentally
more tractable since adding a global loss will lead to a decay of all other modes while the localized defect mode would 
be stable having a constant amplitude.

{\it CSC at the phase transition- } The existence of the delocalization-localization phase transition posses intriguing questions, 
one of which is the nature of the FB mode at the transition point associated with $\epsilon_{cr}^{(I)}$. In particular, it is known 
from the Anderson localization theory, that the eigenfunctions at the metal-to-insulator phase transition are multifractals
i.e. display strong fluctuations on all length scales~\cite{M00,FE95,W80}. Their structure is quantified by analyzing the dependence 
of their moments ${\cal I}_p$ with the system size $N$:
\begin{equation}
\label{PN}
{\cal I}_p = \frac {\sum_n \left|\psi_n\right|^{2p} }{(\sum_n \left|\psi_n\right|^{2})^2}\propto N^{-(p-1)D_{p}}.
\end{equation}
Above the multifractal dimensions $D_p\neq 0$ are different from the dimensionality of the embedded space $d$. Among all 
moments, the so-called inverse participation number (IPN) ${\cal I}_2$ plays the most prominent role. It can be shown that it is roughly 
equal to the inverse number of non-zero eigenfunction components, and therefore is a widely accepted measure to characterize the
extension of a state. We will concentrate our analysis on ${\cal I}_2$ of the FB mode at the phase transition point $\epsilon_{cr}^{(I)}$.

We assume that the eigenmodes of Eq.~(\ref{eq3}) take the following form:
\begin{equation}
\begin{split}
\phi_{n}^{(k)}=A^{(-)}e^{iq^{(k)}n}+B^{(-)}e^{-iq^{(k)}n} (n<0)\\
\phi_{n}^{(k)}=A^{(+)}e^{iq^{(k)}n}+B^{(+)}e^{-iq^{(k)}n} (n>0)\\
\end{split}
\label{eq7}
\end{equation}
where $q^{(k)}=q_{r}^{(k)}+iq_{i}^{(k)}$, while the associated propagation constants are in general complex and can be written in 
the form $\beta^{(k)}\equiv -2V\cos(q^{(k)})=\beta_{r}^{(k)}+i\beta_{i}^{(k)}$. Imposing hard wall boundary conditions to the solutions 
Eq.~(\ref{eq7}) i.e. $\phi_{M+1}=\phi_{-M-1}=0$ allow us to express the coefficients $B^{(-)}, B^{(+)}$ in terms of $A^{(-)}, A^{(+)}$: 
\begin{equation}
B^{(\mp)}=-A^{(\mp)}e^{\mp 2iq(M+1)}
\label{eq8}
\end{equation}
At the same time the requirement for continuity of the wavefunction at $n=0$ lead us to the relation
\begin{equation}
\label{eq9}
A^{(+)}+B^{(+)}=A^{(-)}+B^{(-)}
\end{equation}
Substitution of Eqs.~(\ref{eq8},\ref{eq9}) back into Eq.~(\ref{eq3}) for $n=0$, lead to a transcendental equation for $q$:
\begin{equation}
V\sin[2(M+1)q] \sin(q) = i \epsilon_0^{(I)} \sin^2[(M+1)q]
\label{eq10}
\end{equation}
which can be re-written in terms of two equations
\begin{equation}
\sin[(M+1)q]=0;\,\,{\rm or}\,\,  \cot[(M+1)q] \sin(q)=i{\epsilon_0^{(I)}\over 2V}
\label{eq10b}
\end{equation}
We are interested in the structure of the FB mode in the middle of the band corresponding to ${\cal R}e(\beta)=0$. For 
simplicity of the calculations we assume below that $M+1$ is odd \cite{note2} and also remind that the total size of the system 
is N=2M+1. Imposing the condition ${\cal R}e(\beta)=0$ in the second term of the  Eq.~(\ref{eq10b}) we get that $q_r=-s\pi/2$ 
while the imaginary part $q_i$ satisfies the following equation
\begin{equation}
\label{eq11}
s\tanh\bigg[(M+1)q_i\bigg] \cosh(q_i)=\frac{\epsilon_0^{(I)}}{2V}
\end{equation}
We will look for a stationary solution at the phase transition point $\epsilon_0^{(I)}=s\, 2 V$ with $\beta_{i}\rightarrow 0$ 
(or equivalently $q_{i}\rightarrow 0$) in $N\rightarrow\infty$ limit that also satisfies $q_i\times(M+1)\sim q_i N\rightarrow\infty$ 
condition. In Eq.~(\ref{eq11}) we now perform small $q_i$ expansion in $\cosh(q_i)\approx 1+q_i^2/2$ and large $(M+1)q_i$ 
expansion in $\tanh[(M+1)q_i]=\frac {\exp((M+1)q_i)-\exp(-(M+1)q_i)}{\exp((M+1)q_i)+\exp(-(M+1)q_i)} \approx 1-2\exp(-(N+1)q_i)
\approx 1-q_i^2/2$. In the large $M$-limit the solution of the last transcendental equation can be found by using the definition 
of Lambert W-function. We have 
\begin{equation}
q_i\sim2\frac{\ln(N+1)}{N+1}.
\label{eq12}
\end{equation}
Substituting back to the expression for the propagation constant we get $\beta=-2V\cos(-s\pi/2+iq_i)\approx -s\,2V\,iq_i$ 
which in the large $N (M)$-limit results in $\beta=0$. Finally, substituting Eqs.~(\ref{eq8},\ref{eq12}) back to Eq.~(\ref{eq7}) 
we get that the corresponding FB mode takes the form  
\begin{equation}
\label{eq13}
\phi_n\propto \exp\left[i(-s\pi/2+iq_i)|n|\right]=\frac{(-s\,i)^{|n|}}{\left(N+1\right)^{2|n|/(N+1)}}
\end{equation}
The FB state described by Eq.~(\ref{eq13}) is not exponentially localized neither it is extended. It rather falls to an exotic family of 
critical states and it can quantify better via the IPN ${\cal I}_2$. Using Eq.~(\ref{PN}) for $p=2$ it is easy to show that the IPN of the 
FB mode of Eq.~(\ref{eq13}) is given by Eq.~(\ref{eq1}). Furthermore, this scaling relation is not consistent with the standard power 
law Eq.~(\ref{PN}) characterizing self-similar (fractal) states. Rather we have an unusual situation of {\it a critical state that it is at 
the verge of being fractal}. To our knowledge such anomalous scaling has been discussed only in completely different context of 
Hermitian random matrix models~\cite{ORC11} and were never found to be present in any physical system. Thus our simple set-up constitutes 
the first paradigmatic system where these CSC can be observed. We have also checked that the critical nature of the defect state is 
not a consequence of the degenerate band-edge \cite{DELA03} being present in the case of the tight-binding system of Eq. (\ref{eq3}). 
This can be achieved by introducing an on-site potential $\epsilon_n^{(R)} =\epsilon_0^{(R)} (-1)^n$ which removes the degeneracy 
at $\beta=0$.  We found that the defect state still exists but its energy ${\cal R}e(\beta)$ is no longer at $0$. It is inside one of two 
bands and away from all band edges. Amazingly, there is still a critical point (which depends on $\epsilon_0^{(R)}$) when the defect 
state becomes critical and ${\cal I}_2\sim \log(N+1)/(N+1)$.

\begin{figure}[tbp]
\begin{center}
\includegraphics[width=3.3in]{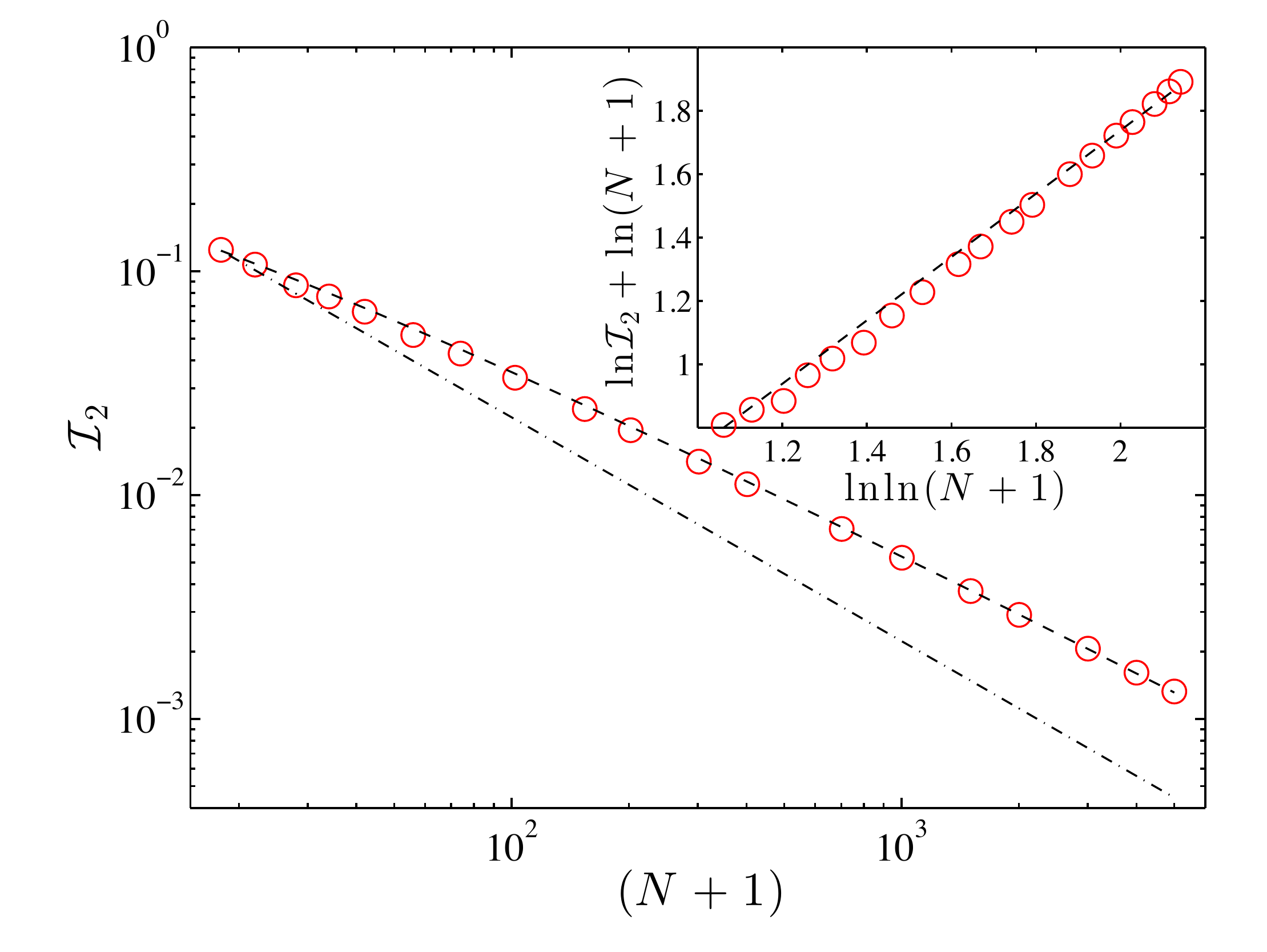}
\end{center}
\vskip -0.5cm
\caption{
Scaling analysis of the IPN ${\cal I}_2$ (see Eq.~(\ref{PN})) of a CSC state versus the system size $N$. 
In the main panel ${\cal I}_2$ (shown as symbols) is plotted versus the size of the system in a double 
logarithmic plot. Dot-dashed line corresponds to $N^{-1}$ dependence. Dashed line is prediction of 
Eq.~(\ref{eq1}) which contains a logarithmic correction. In the inset we plot the same data in a different 
fashion i.e. $\ln{\cal I}_2+\ln(N+1)$ versus $\ln\ln(N+1)$. Straight line with a unit slope confirms the 
existence of the logarithmic dependence as indicated by our theoretical prediction Eq.~(\ref{eq1}). 
}
\label{fig3}
\end{figure}

The validity of our analysis has been confirmed by performing detailed numerical calculations. In the inset of Fig. \ref{fig3} we report 
using a double-logarithmic plot the scaling of ${\cal I}_2$ versus the system size at the phase transition point.  A deviation from a 
straight line (which would be the case of fractal states) is clearly visible as $M$ takes larger values. Instead in the main plot we report 
the anomalous part of ${\cal I}_2$ as a function of $\ln\ln(N+1)$. We see that the data follow a nice straight line, thus confirming the 
validity of our prediction Eq.~(\ref{eq1}).  

{\it Conclusions - } In conclusion we have investigated the structure of non-Hermitian defect states as a function of the 
defect strength. We have found that these states experienced a phase transition from delocalization to localization as 
the imaginary part of the refractive index in the defect waveguide approaches a critical value. At the transition point the 
inverse participation number of this mode scales as $\ln(N)/N$ indicating a weak criticality. This phase transition is 
accompanied by a mode re-organization which reveals analogies with the Dicke super-radiance. The transition survives 
periodic pertubations in the refractive index in the waveguide array and the anomalous logarithmic behavior 
of the inverse participation ratio at the critical point is preserved. It will be interesting to investigate whether this 
behavior survives in higher dimensions and other type of configurations.

\textit{Acknowledgement - } We thank A. Ossipov and Y. Fyodorov for useful discussions. This work was sponsored partly 
by grants NSF ECCS-1128571, DMR-1205223, ECCS-1128542 and DMR-1205307 and by an AFOSR MURI grant FA9550-14-1-0037.

\newpage
\section{Supplemental Material}

In this section, we show that the critical nature of the defect state is not an artifact of degenerate band-edges appearing 
in the middle of the band for the tight-binding model of Eqs. (2,3) of the main text. In order to remove the $\beta=0$ 
degeneracy we introduce a staggering on-site potential as $\epsilon_n^{(R)}$=$\epsilon_0^{(R)}(-1)^n$. Therefore, the 
new tight-binding equation is:

\begin{equation}
\label{eqS1}
\beta^{(k)} \phi_n^{(k)} =- V(\phi_{n+1}^{(k)}+\phi_{n-1}^{(k)}) - (\epsilon_0^{(R)}(-1)^n+\epsilon_n^{(I)}\delta_{n0})\phi_n^{(k)};\\ 
\tag{$S1$}
\end{equation}
We propose the following ansatz for odd/even (denoted by superscript o/e) waveguide numbers:

\begin{equation}
\begin{split}
\phi_{n}^{(k)(o/e)}=A^{(-)(o/e)}e^{iq^{(k)}n}+B^{(-)(o/e)}e^{-iq^{(k)}n} (n<0)\\
\phi_{n}^{(k)(o/e)}=A^{(+)(o/e)}e^{iq^{(k)}n}+B^{(+)(o/e)}e^{-iq^{(k)}n} (n>0)\\
\end{split}
\label{eqS2}
\tag{$S2$}
\end{equation}
In the absence of imaginary defects we get the following dispersion relation:
\begin{equation}
\label{eqS3}
\beta^{(k)}=\pm \sqrt{(\epsilon_0^{(R)})^2+4V^2\cos^2q^{(k)}}
\tag{$S3$}
\end{equation}
Therefore, the degenerate energy at zero is shifted into the positive or negative branch. 

In the presence of defect, and after taking into account the hard wall boundary conditions ($\phi_{M+1}^{(k)(o)}=\phi_{-M-1}^{(k)(o)}=0$) and continuity at n=0, we get two discrete equations for the complex propagation constant $q$:
\begin{equation}
\begin{split}
\sin[(M+1)q]=0;\,\,{\rm or}\,\,\,\,\,\,\,\,\,\,\,\,\,\,\,\,\,\,\,\,\,\,\,\,\,\,\,\,\,\,\,\,\,\,\,\,\,\,\,\,\,\,\,\,\,\,\,\, \\
\cot[(M+1)q] \sin(q)=i{\epsilon_0^{(I)}\over 2V} . \frac{\epsilon_0^{(R)}+\sqrt{(\epsilon_0^{(R)})^2+4V^2\cos^2q} } {2V\cos q}\\
\end{split}
\label{eqS4}
\tag{$S4$}
\end{equation}
The above equations are consistent with the results presented in the main text at the limit $\epsilon_0^{(R)}$$\rightarrow$ 0.

\begin{figure}[tbp]
\begin{center}
\includegraphics[width=3.3in]{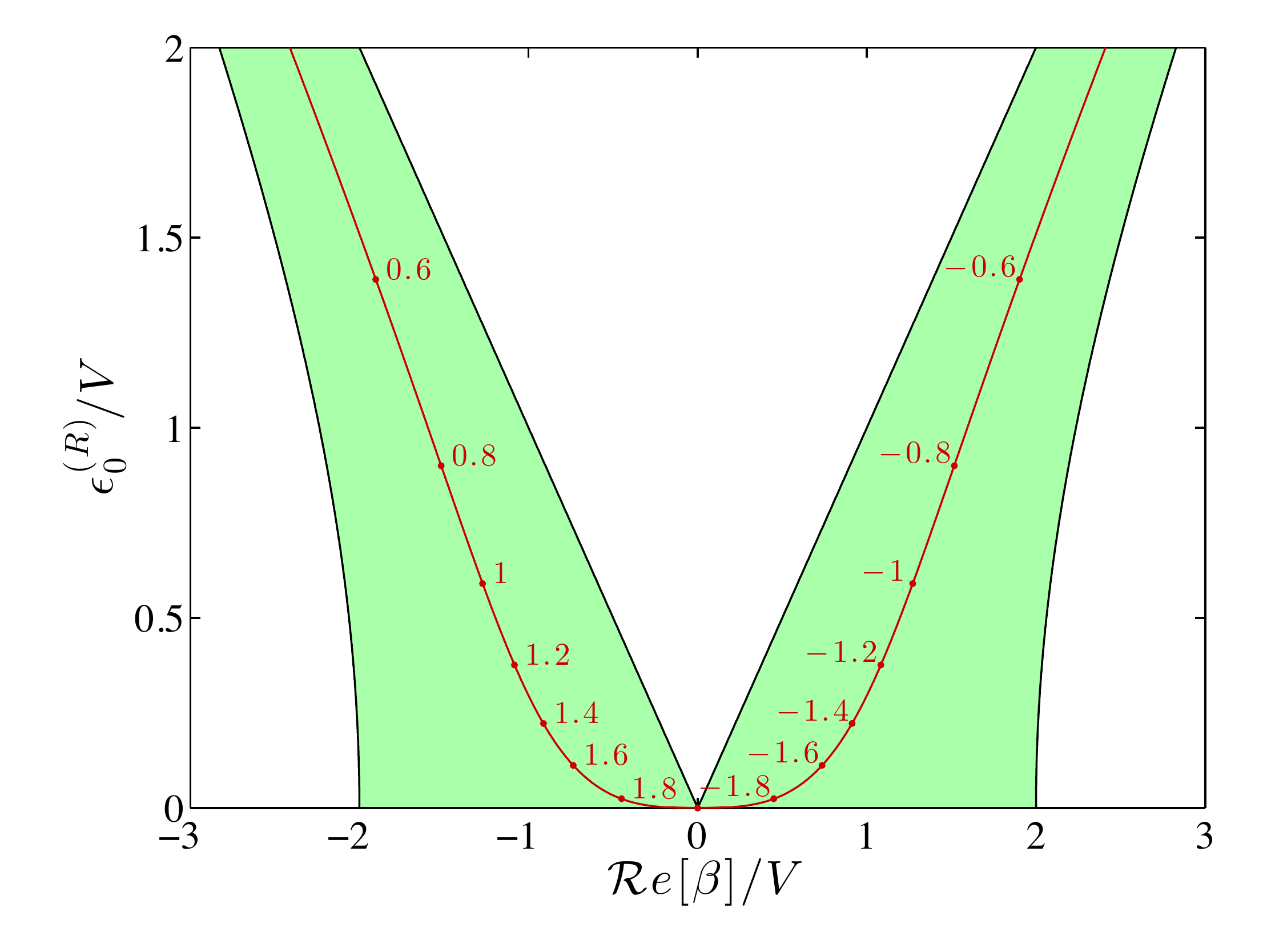}
\end{center}
\vskip -0.5cm
\caption{Band structure (green shadowed array) of the model Eq. (\ref{eqS1}) vs. $\epsilon_0^{(R)}$. The red line indicates the trajectory of the 
defect eigenmode as $\epsilon_0^{(R)}$ increases. The red dots and the associated numbers are indicative values of the critical $\epsilon_0^{(I)}$ 
(for the specific $\epsilon_0^{(R)}$) above which a defect mode is created.
}
\label{fig1}
\end{figure}

In the localized regime ($|\epsilon_0^{(I)}|$$>$ $|\epsilon_{cr}^{(I)}|$), we get $\cot[(M+1)q]\approx i$. By replacing this expression into 
the second term of  Eq.~(\ref{eqS4}), we derive the following cubic relation for $x\equiv\tan q$:
\begin{equation}
\label{eqS42}
\begin{split}
2\epsilon_0^{(R)}\epsilon_{0}^{(I)}x^3+\Big((\epsilon_{0}^{(I)})^2-4V^2\Big)x^2+2\epsilon_0^{(R)}\epsilon_{0}^{(I)}x+(\epsilon_{0}^{(I)})^2=0
\end{split}
\tag{$S5$}
\end{equation}
The above algebraic equation has three roots. Depending on the value of $\epsilon_0^{(I)}$ these roots can be either real or complex. In the 
former case (i.e. $x$, and therefore $q$, being real) the associated mode is extended, while in the latter one (i.e. $x$, and therefore 
$q$, being complex) the associated mode is localized. The transition between these types of modes occurs at $\epsilon_{cr}^{(I)}$ and
is given as a solution of the following equation:
\begin{equation}
\label{eqS5}
\begin{split}
 \bigg((4V^2-\epsilon_{cr}^{(I)})^2\bigg)^3=\,\,\,\,\,\,\,\,\,\,\,\,\,\,\,\,\,\,\,\,\,\,\,\,\,\,\,\,\,\,\,\,\,\,\,
\,\,\,\,\,\,\,\,\,\,\,\, \,\,\,\,\\
8(\epsilon_0^{(R)})^2\bigg(-2V^4+10V^2(\epsilon_{cr}^{(I)})^2+(\epsilon_{cr}^{(I)})^4+
2(\epsilon_{cr}^{(I)})^2(\epsilon_0^{(R)})^2\bigg)
\\
\end{split}
\tag{$S6$}
\end{equation}
Furthermore, it can readily be confirmed that, as expected, for $\epsilon_0^{(R)}\rightarrow 0$, $\epsilon_{cr}^{(I)}$ approaches to 2.

The associated energy $\beta_{cr}$ 
of the defect (localized) mode is found after substituting the expression for $\epsilon_{cr}^{(I)}$ from Eq. (\ref{eqS5}), into Eq. (\ref{eqS4}). This allows us to
evaluate $q^{(cr)}$ which can then be substituted in Eq. (\ref{eqS3}) in order to get an expression for $\beta_{cr}$.

Next, we investigate the scaling behavior of the defect mode at the transition point $\epsilon_{cr}^{(I)}$.  Following the same argumentation as 
used in the main text, we write $q^{(cr)}$ as $q_r^{(cr)}+i q_i$, where we assume that $(M+1) q_i \rightarrow \infty$ and $q_i$ is
a small quantity. Substituting back to the 
transcendental equality of Eq.~(\ref{eqS4}) and expanding each term up to first order in $q_i$ we eventually get:
\begin{equation}
\label{eqS6}
q_i\sim\frac{\ln(N+1)}{N+1}.
\tag{$S7$}
\end{equation}
Considering the fact that ${\cal I}_2\sim q_i$, it can be deduced that the second moment of the defect mode for the modified model scales anomalously as indicated in
Eq. (1) of the main text.

To summarize, we elucidated that the logarithmic scaling of IPR is not a consequence of degenerate band-edge in Anderson model at $\beta=0$.


\begin{thebibliography}{99}
\bibitem{P93} A. Peres, {\it Quantum Theory: Concepts and Methods}, Kluwer Academic Publishers (1993).

\bibitem{M67} N. F. Mott, Adv. Phys. {\bf 16}, 49 (1967).

\bibitem{NW29} J. von Neumann and E. Wigner, Z. Phys. {\bf 30}, 465 (1929).

\bibitem{PPDHNSS11} Y. Plotnik {\it et al}, Phys. Rev. Lett. {\bf 107}, 183901 (2011).

\bibitem{WXKMTNSSK13} S. Weimann {\it et al}, Phys. Rev. Lett. {\bf 111}, 240403 (2013).

\bibitem{CVCOL13} G. Corrielli {\it et al}, Phys. Rev. Lett. {\bf 111}, 220403 (2013).

\bibitem{HZLCJJS13} C. W. Hsu {\it et al}, Nature {\bf 499}, 188 (2013).

\bibitem{PE05} R. Porter, D. Evans, Wave Motion {\bf 43}, 29 (2005); C. M. Linton, P. McIver, Wave Motion {\bf 45}, 16 (2007).

\bibitem{CSFSCC92}F. Capasso, {\it et al}, Nature{\bf 358}, 565 (1992).

\bibitem{MBS08} D. C. Marinica, A. G. Borisov, S. V. Shabanov, Phys. Rev. Lett. {\bf 100}, 183902 (2008).

\bibitem{OPR03} J. Okolowicz, M. Ploszajczak, I. Rotter, Phys. Rep. {\bf 374}, 271 (2003).

\bibitem{ZGWL10}K. Zhou {\it et al}, Opt. Lett. {\bf 35}, 2928 (2010)

\bibitem{RMBNOCP13}A. Regensburger {\it et al}, Phys. Rev. Lett. {\bf 110}, 223902 (2013)

\bibitem{L14} S. Longhi, {\it Bound states in the continuum in ${\cal PT}$-symmetric optical lattices}, arXiv:1402.3761 (2014)

\bibitem{CLS03} D.N. Christodoulides, F. Lederer, and Y. Silberberg, Nature {\bf 424}, 817 (2003).

\bibitem{note1} It is possible to have the same $\epsilon_0^{(R)}$ for the defect waveguide (without violating the Kramers-Kronig relations).
One way to achieve this is by correcting the changes in the $\epsilon_0^{(R)}$ at $n=0$, due to the presence of $\epsilon_0^{(I)}$, by 
appropriate adjustment of its width.

\bibitem{salamo} A. Guo, {\it et. al.}, Phys. Rev. Lett. {\bf 103}, 093902 (2009).

\bibitem{szameit} T. Eichelkraut {\it et al}, Nature Communcations {\bf 4}, 2533 (2013)

\bibitem{kipp} C. E. Ruter {\it et. al}, Nat. Phys. {\bf 6}, 192 (2010).

\bibitem{D54}R.H. Dicke, Phys. Rev. {\bf 93}, 99 (1954).

\bibitem{SZ89} V. V. Sokolov, V. G. Zelevinsky, Nucl. Phys. {\bf A504}, 562 (1989).

\bibitem{C12} G. L. Celardo {\it et al.}, J. Phys. Chem. C {\bf 116}, 22105 (2012); R. Monshouwer {\it et al}, J. Phys. Chem. B {\bf 101}, 7241 (1997).

\bibitem{KSKHSA12} J. Keaveney {\it et al}, Phys. Rev. Lett. {\bf 108}, 173601 (2012); M. O. Scully, A. A. Svidzinsky, Science {\bf 328}, 1239 (2010).

\bibitem{E06} E. N. Economou, {\it Green's Functions in Quantum Physics}, Springer Series in Solid-State Sciences (Third Edition)  (2006).

\bibitem{M00} A. D. Mirlin, Phys. Rep. {\bf 326}, 259 (2000); Y. V.~Fyodorov and A. D.~Mirlin,
Int. J. Mod. Phys. {\bf 8}, 3795 (1994); Y. V. Fyodorov and A. D. Mirlin, Phys. Rev. B
{\bf 51}, 13403 (1995).

\bibitem{FE95}V. I. Falko and K. B. Efetov, Europhys. Lett. {\bf 32}, 627 (1995);
Phys. Rev. B {\bf 52}, 17413 (1995).

\bibitem{W80} F. Wegner, Z. Phys. B {\bf 36}, 209 (1980); H. Aoki, J. Phys. C {\bf 16},
L205 (1983); M. Schreiber and H. Grussbach, Phys. Rev. Lett. {\bf 67}, 607 (1991); D. A.
Parshin and H. R. Schober, ibid. {\bf 83}, 4590 (1999); A. Mildenberger, F. Evers, and 
A. D. Mirlin, Phys. Rev. B {\bf 66}, 033109 (2002). 

\bibitem{note2} We note that for even $N$ there are two critical states that emerge symmetrically around ${\cal R}e(\beta)=0$. The rest of the analysis
remains qualitatively the same.

\bibitem{ORC11}A. Ossipov, I. Rushkin, and E. Cuevas, Journal of Physics: Cond. Matt. {\bf 23}, 415601 (2011).

\bibitem{DELA03}L.I. Deych {\it et al}, Phys. Rev. Lett. {\bf 91}, 096601 (2003)

\end{thebibliography}
\end{document}